\documentclass[twocolumn]{article}

\title{\bf Magnetic field and radius of innermost stable circular orbit near SMBH in AGNs}
\author{M.Yu. Piotrovich\thanks{E-mail:mike@gao.spb.ru}, Yu.N. Gnedin\thanks{E-mail:gnedin@gao.spb.ru}, N.A. Silant'ev, T.M. Natsvlishvili and S.D. Buliga\\
Central Astronomical Observatory at Pulkovo of Russian Academy of Sciences, \\Pulkovskoye chaussee 65, Saint-Petersburg, 196140, Russia}

\begin{document}

\maketitle

\begin{abstract}
Magnetic fields in an accretion disk around the central black hole can modify the position of the innermost stable circular orbit (ISCO) radius and produce the difference from the classical Novikov-Thorne radius. We estimated the ISCO magnetic field strength on the base of polarimetric observations of the accretion disk radiation. This estimate can be obtained with taking into account the effect of Faraday rotation of the polarization plane at the length of the free path of photon between successive electron scattering events. In a result we presented the new method for real estimation of the ISCO radius in the accretion disk, i.e. in the nearest vicinity of a central black hole. Our estimations confirmed the \cite{frolov14} conclusion that magnetic field produces the effect in a result of which the innermost stable circular orbit becomes closer to the horizon of a black hole.
\end{abstract}


\section{Introduction} 

Recently, \cite{frolov14} have discussed the interesting aspect of spectral line broadening, namely broadening of iron $K_{\alpha}$ lines, as probes of the magnetic field in the black hole vicinity. They have underlined that in the presence of a regular magnetic field in the vicinity of black holes the characteristic of the motion of charged ions are modified. In a result the situation appears that their innermost stable circular orbits (ISCO) become close to the BH horizon. This effect modifies the spectrum broadening of lines emitted by such an iron ion.

Recently \cite{shaymatov14} claimed that magnetic field can serve as a cosmic censor of Kerr black hole. They confirmed that magnetic field although small, affects motion of charged particles drastically due to the large Lorentz force and this electromagnetic force can be much stronger that the gravity.

There are both theoretical and observational evidence that magnetic field plays an important role in the black hole physics. It is well known that the magnetic field is essential to angular momentum transfer in accretion disks \cite{balbus98,krolik01,pariev03}. In the Blandford-Znajek model a regular magnetic field is required to explain black hole jets energies \cite{blandford77}. \cite{frolov14} have demonstrated that the position of the ISCO for a charged particles is close to the black hole horizon radius that standard ISCO radius for Schwarzschild black hole: $R_{ISCO} = 6 R_g$. Here $R_g = G M_{BH} / c^2$ and $M_{BH}$ is the black hole mass, $G$ is the gravitational constant and $c$ is the light speed.

For investigation of this situation we use our method of determination of ISCO magnetic field strength based on polarimetric observation of the accretion flow radiation. The base of this method consists in taking into account the effect of Faraday rotation of the polarization plane and Faraday depolarization at the length of the free path of photon between successive electron scattering events \cite{gnedin84,gnedin97,dolginov95,silantev09}. In a result the difference between the standard \cite{chandra50} and observed polarization degree allows to estimate the magnitude of magnetic field in the radiation region \cite{silantev09}. Then using the effective model of an accretion disk it is possible to estimate the ISCO magnetic field strength. The ISCO magnetic field is the function of $R_{ISCO}$, i.e. the function of the ISCO radius itself: $B_{in} \sim f(q)$, where $q = R_{ISCO} / R_g$. Then using well known relation between the relativistic jet kinetic power $L_j$ and the spin of a SMBH one obtains the following expression for the jet kinetic power:

\begin{equation}
 L_j = \left(B_{ISCO} / 10^5 G \right)^2 M_9^2 f(a),
 \label{eq01}
\end{equation}

\noindent where $M_9 = M_{BH} / 10^9 M_{\odot}$, $M_{BH}$ is the SMBH mass, $a$ is the SMBH spin and function $f(a)$ is tabulated in \cite{garofalo10}. Later on we shall use the expression for $f(a)$, presented for the tandem of Blandford-Znajek and Blandford-Payne physical processes of generation of relativistic jets in AGN (Fig.3 from \cite{garofalo10}).

\section{The basic relations} 

For determination of the magnetic field strength for the ISCO ($B_{ISCO}$) we used the results of polarimetric observations of broad line region (BLR) emission in AGN \cite{smith02} and the polarimatric data obtained at BTA-6m telescope of Special Astrophysical Observatory of RAS \cite{afanasiev2011}.

The gas velocities in the BLR size are usually used to obtain an estimate of the central black hole mass. Most of the recent studies present the conclusion that the BLR is a flattened rotating system \cite{punsly10,kollatschny13,tremaine13}. For example, \cite{kollatschny13} have modeled the rotational and turbulent velocities in BLR and estimated of the ratio of the characteristic thickness of the disk to the radius as $H / R > 0.2$. It means that the BLR can be considered as the geometrically and optically thick disk.

This fact allows us to use the results of \cite{silantev13} paper for calculations of polarization degree and its dependence on the magnetic field strength $B(R_{BLR})$ in BLR.

\begin{equation}
 P_l(R_{BLR}) = \frac{P_L(\mu)}{\sqrt{1 + b^2}},
 \label{eq02}
\end{equation}

\noindent where $\mu = \cos{i}$ and $i$ is the disk inclination angle. The polarization degree $P_l(\mu)$ is corresponding to the solution of the classical Milne problem in non-magnetized plane parallel atmosphere \cite{chandra50}.

The dimensional parameter $b$ describes the effect of Faraday depolarization:

\begin{equation}
 b = 0.8 \lambda^2 \sqrt{1 - \mu^2} B_{\perp}(R),
 \label{eq03}
\end{equation}

\noindent where $B_{\perp}(R) = \sqrt{B_{\rho}^2 + B_{\varphi}^2}$ is the magnetic field component that is parallel to the accretion disk plane. The quantity $0.8 \lambda^2 B$ is equal to the Faraday rotation angle of the polarization plane at the Thomson optical depth of $\tau = 2$ if the polarized radiation propagates along the magnetic field direction. The magnetic field is measured in Gauss and the wavelength is measured in $\mu m$. The wavelength $\lambda$ corresponds to the rest frame system.

The equations (\ref{eq02}) and (\ref{eq03}) allow to derive the magnetic field strength at the distance $R$ in the accretion flow from the central gravitating source.

We use the BLR parameters for estimation of magnetic field $B(R_{BLR})$, where $R_{BLR}$ is the radius of the BLR in AGNs. The results of polarimetric observations of BLR emission for the AGN sample are presented by \cite{smith02}. The relations (\ref{eq02}) and (\ref{eq03}) allow us to derive the magnetic field strength $B(R_{BLR})$. For estimation of the ISCO magnetic field strength we use the following equation:

\begin{equation}
 B(R_{ISCO}) = B(R_{BLR}) \frac{R_{BLR}}{R_{ISCO}}.
 \label{eq04}
\end{equation}

The simplest and best studied magnetic configuration is most likely toroidal one \cite{bonanno07}, i.e. $B_{\perp} \gg B_z$. A contributing factor is the differential keplerian rotation in the accretion disk, which transforms the poloidal magnetic field into a toroidal one \cite{bonanno07}.

Black hole masses in AGNs are derived using the BLR gas as the dynamical tracer, based on the assumption that the gas is primarily accelerated by the gravity of the black hole. The gas velocity dispersion is derived from the broad line width, but the black hole mass estimation itself requires also the radius of the BLR. The best estimation for the BLR size comes from ''reverberation'' or ''echo'' mapping \cite{blandford82,greene10}. Using the best estimation, \cite{greene10} presented the $R_{BLR}$ values for the sample of AGN (Table 1 from \cite{greene10}) and we used these data. Besides \cite{shen10} proposed the following empirical analytical formula for $R_{BLR}$ that is quite well corresponding to the $R_{BLR}$ data from ''reverberation'' method:

\begin{equation}
 R_{BLR} = 2.1 \times 10^{17} M_8^{1/2} l_E^{1/2} cm.
 \label{eq05}
\end{equation}

We shall use both possibilities for determination of BLR radius: ''reverberation'' method \cite{greene10} and theoretical estimation (\ref{eq05}). We demonstrate below how to use the equations (\ref{eq01})-(\ref{eq05}) one can determine the $R_{ISCO}$ or the corresponding parameter $q = R_{ISCO} / R_g$.

\section{Determination the ISCO radius in accretion disk around the central black hole in Fairall~9}

Fairall~9 galaxy is classified as Sy 1,2. The current status of X-ray reflection provides black hole spin measurements. According to \cite{reynolds13} the spin value of SMBH in Fairall~9 is $a = 0.52_{-0.15}^{+0.19}$. \cite{brenneman11} give the value $a = 0.65 \pm 0.5$, that does not contradict to \cite{reynolds13} value. The mass of Fairall~9 SMBH is $\log{M_{BH}/M_{\odot}} = 8.41 \pm 0.11$ and Eddington ratio $l_E = L_{bol} / L_{Edd} = 0.05 \pm 0.01$ \cite{Brenneman13}. The position angle $i = {45^{\circ}}_{-9}^{+13}$ is determined in \cite{walton10}. \cite{lohflink12} give higher value of the spin: $a = 0.71_{-0.09}^{+0.08}$ and the little lower inclination: $i = {37^{\circ}}_{-2}^{+4}$. Estimation of BLR radius from (\ref{eq05}) gives $R_{BLR} = 7.5 \times 10^{16} cm$. From ''reverberation'' method \cite{greene10} we have $R_{BLR} = (17.4 \pm 1.3)$ lt. days. $R_{BLR}$ maximal value from the ''reverberation'' method is $R_{BLR}(max) = 4.85 \times 10^{16} cm$.

For estimation of $B_{ISCO}$ we use the Eq.(\ref{eq01}). According to \cite{merloni07} the kinetic power $L_j$ of the relativistic jet can be derive as

\begin{equation}
 \log{L_j / L_{Edd}} = (0.49 \pm 0.07) \log{l_E} - (0.78 \pm 0.36)
 \label{eq06}
\end{equation}

The next step is to determine the ISCO magnetic field using the data of spectropolarimetric observations from \cite{smith02}. They derived the following value for the degree of polarization in BLR: $P_l(obs) = 0.37 \pm 0.13 \%$. Our Eq.(\ref{eq02}) and Eq.(\ref{eq03}) allow to estimate the strength of the magnetic field in BLR as $B_{\perp}(R_{BLR}) = 9.3 G$. Then using the relation (\ref{eq04}), one obtains $B_{ISCO} \equiv B(R_{ISCO}) = 10^4 \times 1.1 / q(a) G$. We use for this estimation the value of the inclination angle obtained by \cite{walton10}, i.e. $i = 45^{\circ}$. The Eq.(\ref{eq06}) gives the next value for the jet kinetic power: $L_j = 10^{45.1} erg/s$. In a result for the spin value $a = 0.5$ the Eq.(\ref{eq01}) gives the following value for the dimensionless ISCO radius: $q = 2.24$. But in the classical work \cite{novikov73} the dimensionless ISCO radius is $q = 4.23$ \cite{aliev12}. In confirms really the effect of presence of magnetic field in the vicinity of ISCO producing phenomenon in which the BH ISCO becomes closer to the BH horizon \cite{frolov14}.

The results of our calculations for SMBH in other AGNs are presented in the Table 1.

\begin{table}[tbp]
\centering
\begin{tabular}{|l|l|l|l|}
\hline
AGN         & $a$  & $q_{MF}$   & $q_{NT}$ \\
\hline
Mrk 335     & 0.7  & 1.7        & 3.39 \\
Fairall 9   & 0.5  & 2.24       & 4.23 \\
Ark 120     & 0.64 & 2.6        & 3.83 \\
Ark 564     & 0.96 & 1.1        & 1.6  \\
Mrk 509     & 0.4  & 2.34       & 4.61 \\
Mrk 841     & 0.5  & $\geq 1.0$ & 3.83 \\
NGC 4051    & 0.5  & 3.47       & 4.23 \\
NGC 3783    & 0.9  & 1.57       & 2.32 \\
I Zw I      & 0.55 & 2.7        & 4.0 \\
PG 1244+026 & 0.9  & 1.65       & 2.3 \\
NGC 7469    & 0.7  & 1.8        & 3.39 \\
NGC 5548    & 0.9  & 1.2        & 2.32\\
\hline
\end{tabular}
\caption{\label{tab:i} The results of our calculations for some AGNs. $a$ is the dimensionless spin, $q_{MF} = R_{ISCO} / R_g$, $q_{NT}$ is a standard dimensionless ISCO radius.}
\end{table}

\section{Comments to the Table 1}

The Table 1 presents the list of AGNs with determined spins \cite{Brenneman13,reynolds13,emmanoulopoulos14,sesana14,crummy06}.

For estimations of standard values of the ISCO radius we used the results of \cite{aliev12}. We used the polarimetric data from \cite{smith02}. Data about BLR radius $R_{BLR}$ are presented by \cite{greene10}.

The basic problem is determination the inclination angle of an accretion disk in AGNs. Namely the value of the inclination angle determined the standard (Sobolev-Chandrasekhar) magnitude of the linear polarization $P_l(\mu)$. The magnitudes of inclination angles for our objects from Table 1 are presented in publications \cite{crummy06,ho08,emmanoulopoulos14,walton10,wu01}.

\section{Conclusion}

We estimated the ISCO magnetic field strength on the base of polarimetric observations of the accretion disk radiation taking into account the effect of Faraday rotation of the polarization plane and Faraday depolarization effect at the length of the free path of photon between successive electron scattering events. It allows to estimate the magnitude of a magnetic field in the radiation region.Then using the accretion disk model it is possible to estimate the ISCO magnetic field strength that is the function of the ISCO radius $R_{ISCO}$. In a result we obtained the new method for the real estimation of the radius of the ISCO $R_{ISCO}$ in the accretion disk, i.e. in the nearest vicinity of a central black hole. Thus we obtain the method allowing to test the hypothesis of \cite{frolov14} that magnetic field modifies the position of $R_{ISCO}$ and produces the difference on the classical Novikov-Thorne value. This phenomenon allows the magnetic field to be a cosmic censor of Kerr black holes as it is formulated in \cite{shaymatov14}.

\section*{Acknowledgments}

This work was supported by the Basic Research Program of the Presidium of the Russian Academy of Sciences P-41 and the Basic Research Program of the Division of Physical Sciences of the Russian Academy of Sciences OFN-17.

\end{document}